# Why $\vec{p} = \gamma(v) m\vec{v}$ instead of $\vec{p} = m\vec{v}$? Because of the relativistic postulate

**Bernhard Rothenstein**

Politehnica University of Timisoara, Physics Dept., Piata Horatiu 1, Timisoara, Romania

E-mail: bernhard_rothenstein@yahoo.com

**Corina Nafornita**

Politehnica University of Timisoara, Communications Dept., Timisoara, Romania

**Abstract**
It is shown that the addition law of relativistic velocities leads to the fundamental equations of relativistic dynamics.

Chrysos[1] considers that the introduction of the fundamental relationship $\vec{p} = \gamma(v) m\vec{v}$ in a few hours course on special relativity for beginners is a delicate task. He derives it "economically" starting with an useful formula which relates the velocities $\vec{v}(v_x, v_y)$ and $\vec{v}'(v'_x, v'_y)$ of the same particle relative to the reference frames R and R', the latter moving with velocity V in the positive direction of the overlapped axes OX and O'X' of R and R' respectively which are in the standard arrangement. In the best tradition of teaching relativistic dynamics, the Author invokes collisions, conservation of momentum, and mass.

The purpose of our Note is to derive transformation equations for momentum, mass and energy in which we use the addition law of relativistic velocities

$$v_x = \frac{v'_x + V}{1 + \frac{v'_x V}{c^2}} \tag{1}$$

$$v_y = \gamma^{-1} \frac{v'_y}{1+\frac{v'_x V}{c^2}}, \qquad (2)$$

and the relativistic postulate, but avoiding collisions.

The physicist who has learned physics in the R frame considers that mass $m$ and momentum $\vec{p}(p_x, p_y)$ are related by

$$p_x = m v_x \qquad (3)$$

$$p_y = m v_y. \qquad (4)$$

In accordance with the relativistic, observers from R' consider that the mass $m'$ and the momentum $\vec{p}'(p'_x, p'_y)$ are related by

$$p'_x = m' v'_x \qquad (5)$$

$$p'_y = m' v'_y. \qquad (6)$$

Combining Eqs.(1), (3) and (5) we obtain an equation

$$\frac{p_x}{m} = \frac{p'_x}{m'} \frac{1+\frac{V}{v'_x}}{1+\frac{V v'_x}{c^2}} \qquad (7)$$

which suggests that

$$p_x = f p'_x \left(1+\frac{V}{v'_x}\right) \qquad (8)$$

$$m = f m' \left(1+\frac{V v'_x}{c^2}\right) \qquad (9)$$

where $f$ represents a function we should determine.

The inverse transformation equation of Eq.(9) is given by

$$m' = f m (1 - \frac{V v_x}{c^2}) \qquad (10)$$

Multiplying Eqs.(8) and (9) side by side we obtain

$$f = \frac{1}{1-\frac{V^2}{c^2}} = \gamma(V) \qquad (11)$$

and so

$$m = \gamma(V) m' \left(1+\frac{V v'_x}{c^2}\right) = \gamma(V) \left(m' + \frac{V p'_x}{c^2}\right) \qquad (12)$$

$$p_x = \gamma(V) p'_x \left(1+\frac{V}{v'_x}\right) = \gamma(V)(p'_x + m'V). \qquad (13)$$

Combining Eqs.(2), (4) and (6) we obtain

$$\frac{p_y}{m} = \frac{p'_y}{m'} \frac{v_y}{v'_y} = p'_y \qquad (14)$$

the OY(O'Y') component of the momentum being a relativistic invariant.

Multiplying both sides of Eq.(12) by $c^2$ it becomes
$$mc^2 = \gamma(V)(mc^2 + Vp'_x). \tag{15}$$
We have gained that way a new physical quantity $E = mc^2$ in R and $E' = m'c^2$ in R' having the physical dimensions of energy, with which, Eq.(15) becomes
$$E = \gamma(V)(E' + Vp'_x) \tag{16}$$
whereas Eq.(13) can be presented as
$$p_x = \gamma(V)\left(p'_x + V\frac{E'}{c^2}\right). \tag{17}$$
A simple collision considered from R and R' respectively and using the transformation equations derived above leads to the conservation of mass (energy) and momentum.

The concepts of rest energy and kinetic energy are a natural consequence of the results obtained so far. If the particle is at rest in R', observers of that frame measure its rest energy $E_0$. Taking into account that the single supplementary energy a free particle can possess is its kinetic energy $E_K$ associated with the fact that the particle moves with constant velocity V relative to K, is given by
$$E_K = E_0[\gamma(V) - 1]. \tag{18}$$
We consider that such a way to present relativistic dynamics is time saving increasing the pedagogical outcome (understanding/time invested). It avoids collisions, the learning and the teaching of which, puts under a hard trial the memory of the teacher and of the learner as well[2].

**References**

[1] M. Chrysos, Eur.J.Phys. 25, L33-L35 (2004)
[2] G.N. Lewis and R.Tolman, Phil.Mag. 18, 510 (1909)